\documentclass[12pt,eqsecnum,floats,aps,amsmath,amssymb,nofootinbib,prd]{article}

\usepackage{authblk}
\usepackage{setspace}
\usepackage{comment}
\usepackage{setspace}
\usepackage{amsmath,amssymb,amsfonts,amsthm,mathrsfs}
\usepackage{graphicx}
\usepackage{enumerate} 
\usepackage[pdftex]{hyperref}

\def\be{\begin{equation}}
\def\ee{\end{equation}}
\def\ba{\begin{eqnarray}}
\def\ea{\end{eqnarray}}
\def\bi{\begin{itemize}}
\def\ei{\end{itemize}}

\def\xh{\hat{x}}

\def\I{\mathcal{I}}
\def\e{\varepsilon}
\def\qh{\hat{q}}

\def\wb{\bar{w}}
\def\zb{\bar{z}}

\def\t{\tau}
\def\H{\mathcal{H}}

\def\fh{f_\H}
\def\xio{\mathring{\xi}}

\def\st{  \text{ST}}
\def\Vh{V_\H}

\def\lh{\lambda_\H}
\def\Gu{G_{U(1)}}
\def\Gst{G_{\st}}

\title{Null to time-like infinity Green's functions for asymptotic symmetries in Minkowski spacetime}
\author{Miguel Campiglia}
\affil{Universidad de la Rep\'ublica, Montevideo, Uruguay}

\begin{document}
\maketitle

\thispagestyle{empty}

\let\oldthefootnote\thefootnote
\renewcommand{\thefootnote}{\fnsymbol{footnote}}
\footnotetext{Email: campi@fisica.edu.uy}
\let\thefootnote\oldthefootnote

\begin{abstract}
We elaborate on the Green's functions that appeared in \cite{cl3,cl4}  when generalizing, from  massless to massive particles,  various equivalences between soft theorems and Ward identities of large gauge symmetries.  We analyze these Green's functions in considerable detail and show that they form a hierarchy of functions which describe `boundary to bulk' propagators for large $U(1)$ gauge parameters, supertranslations and sphere vector fields respectively. As a consistency check we verify that the Green's functions associated to the large diffeomorphisms map the Poincare group at null infinity to the  Poincare group at time-like infinity.

\end{abstract}
\section{Introduction}
In this note we expand on certain Green's functions recently found  in \cite{cl3,cl4} when extending the action of  large gauge symmetries from massless to massive particles. The general ideas behind such studies are as follows.

The works of Strominger and collaborators  \cite{strom1,strom2,stromst,virasoro} established a way  to relate large gauge symmetries at null infinity with  soft theorems appearing in scattering amplitudes. In the Maxwell case,  the large gauge transformations are given by $U(1)$ gauge parameters $\tilde{\lambda}$ that asymptote to non-trivial functions on the sphere at null infinity:
\be
\lim_{r \to \infty} \tilde{\lambda}(r,u,\xh) = \lambda(\xh). \label{lamo}
\ee
Ward identities associated to these large symmetries where shown to be equivalent to Weinberg's soft photon theorem in \cite{stromqed}, for the case where the charged scattering particles are massless. 

In the gravity case, the large gauge transformations are diffeomorphisms generated by two type of vector fields, exhibiting  the following asymptotic value at null infinity:
\ba
\lim_{r \to \infty} \xi_f^a(r,u,\xh) \partial_a & = & f(\xh) \partial_u , \label{st} \\
\lim_{r \to \infty} \xi_V^a(r,u,\xh) \partial_a & = & V^A(\xh) \partial_A +u \alpha(\xh) \partial_u . \label{vecf} 
\ea
The first ones are known as supertranslations and are parameterized by sphere functions $f(\xh)$. The second ones are `generalized rotations' and are parametrized by   arbitrary vector field on the sphere  $V^A(\xh)$ ($\alpha = (D^A V_A)/2$).\footnote{The standard BMS group (see \cite{aareview} for a recent review) arises by restricting attention to vector fields $V^A$ that are (global) conformally Killing on $S^2$. Barnich and Troessaert's `extended' BMS \cite{barnich2} arises by taking $V^A$ to be local (with singularities) conformal Killing vector fields.  The  `generalized' BMS vector fields (\ref{st}), (\ref{vecf})   can be characterized as spacetime vector fields that are asymptotically divergence-free (rather than Killing as in BMS) at null infinity  \cite{cl1} .}

In \cite{stromst} it was shown that Ward identities associated to supertranslations (\ref{st}) are equivalent to Weinberg's soft theorem \cite{weinberg} and  in \cite{cl1,cl2} it was shown that Ward identities associated to sphere vector fields (\ref{vecf}) are equivalent to Cachazo-Strominger (CS) soft theorem \cite{cs}.\footnote{The works \cite{cl1,cl2} where heavily based in the  work \cite{virasoro} where Virasoro Ward Identities associated to the extended BMS group where shown to be implied by CS soft theorem.}  As in the Maxwell case, both cases where restricted to massless scattering particles.

Since soft theorems are valid for both massless and massive particles, it is natural to ask if there is also a symmetry interpretation in the massive case. At first this seems not  possible: A key fact used  in obtaining the Ward identities is that both large gauge transformations and  scattering particles  `live' at the same place: null infinity. On the other hand, scattering massive particles `live' at time-like infinity. How could the above large gauge transformations act on them?

Now, the various scattering amplitude soft theorems we are referring are obtained from perturbative calculations which are typically performed under some gauge-fixing condition. In particular, the expressions that have been used in the above works rest in formulas derived from perturbative calculations in  harmonic gauges. Of course the soft theorems are gauge invariant,  by which we mean invariant under \emph{small} gauge transformations (i.e. gauge transformations that die down at infinity). However, as it will become clear below, the harmonic gauge lives its imprint in the \emph{large} gauge transformation relevant for the discussion of asymptotic symmetries.

In both the Maxwell and gravity cases the harmonic gauge condition leaves `residual' gauge symmetries associated to  parameters that satisfy the wave equation:
\be \label{boxlam}
\begin{array}{lll}
\square \tilde{\lambda} & = & 0 ,\\
\square \xi_f^a & = & 0 ,\\
\square \xi_V^a & = & 0.
\end{array}
\ee
We can then try to  solve these wave equations with  boundary conditions (\ref{lamo}), (\ref{st}), (\ref{vecf}) to get the gauge parameters in spacetime. The asymptotic behaviour of such solutions  at time-like infinity will then tell us how the 
 large gauge transformations act on the scattering massive particles. In practice one obtains Green's functions that directly give the asymptotic time-like infinity value in terms of the value at null-infinity. As shown in \cite{cl3,cl4}, the associated Ward identities are then equivalent to the corresponding  soft theorems. In this way, one brings massive particles on equal footing as massless particles regarding the relation between asymptotic  symmetries and soft theorems.

The purpose of this note is to present in a unified fashion the various Green's functions that appeared in such studies, and provide derivations of their properties.
The presentation will be in the context of flat Minkowski spacetime. It is the hope that suitable notion of asymptotically flat spacetimes at null and time-like infinities will allow to extend these notions to a general context of nonlinear gravity.

The organization of the material is as follows. In section \ref{prelsec} we review the description  of time-like infinity as a unit spacelike hyperboloid $\H$ that was used in \cite{cl3,cl4} (inspired by the description of spatial infinity given by Ashtekar and Romano \cite{romano}). The wave equations satisfied by the gauge parameters translate into elliptic equations on $\H$ satisfied by the (time-like asymptotic value of) gauge parameters, and we review such equations.
In section \ref{greensec} we describe general  scalar Green's functions on $\H$ and  describe the Green's functions relevant for the large gauge transformations. In section \ref{softsec} we describe the relationships between the Green's functions for large gauge transformations and the associated `soft factors' that appeared in the soft theorems. In section \ref{poincare} we make a  check of the formulas and verify that for the Poincare subgroup of the `generalized' BMS group (\ref{st}), (\ref{vecf}) one recovers the expected time-like asymptotic values.

\section{Hyperboloid description of time-like infinity and asymptotic value of gauge parameters} \label{prelsec}
Let $x^\mu=(t, \vec{x})$ be Cartesian coordinates of Minkowski spacetime. In the region $t \geq r \equiv \sqrt{\vec{x} \cdot \vec{x}}$ we introduce `hyperbolic coordinates'  $(\t,\rho,\xh)$:
\be
\t = \sqrt{t^2 - r^2} ,  \quad \quad \rho = \frac{r}{\sqrt{t^2 - r^2}},  \quad \quad  \xh= \vec{x}/r  \label{carttohyp}
\ee
\be
t= \t \sqrt{1+ \rho^2}, \quad \quad  \vec{x} =  \rho \, \t \, \xh . \label{hyptocart}
\ee
Minkowski metric in these coordinates takes the form
\be
ds^2 = - d \t^2 + \t^2  d \sigma^2, \label{metric}
\ee
where
\be
d \sigma^2  =  \frac{d \rho^2}{1+ \rho^2} + \rho^2 \gamma_{AB} d x^A d x^B   \; = : h_{\alpha \beta} d x^\alpha d x^\beta,
\ee
is the  metric of a  unit space-like hyperboloid that we denote by $\H$.  We use letters $A,B,\ldots$ for sphere coordinates and $\alpha,\beta$ for coordinates on the hyperboloid.  $\gamma_{AB}$ is the unit sphere metric  and $h_{\alpha \beta}$ the unit hyperboloid metric. $\H$  provides a manifold structure for time-like infinity, in the same spirit as null infinity is described by the null manifold $\I$.  Whereas asymptotic massless particles have their home at $\I$, asymptotic massive particles of rest mass $m$ and momentum $\vec{p}$ live at $\H$, according to the identification: 
\be
\rho = |\vec{p}|/m, \quad \quad  \xh= \vec{p}/|\vec{p}| .\label{prho}
\ee
Our interest is in describing the time-like infinity asymptotic value of the large gauge parameters described in the introduction.  In \cite{cl3},\cite{cl4} it was argued that such parameters have the  $\t \to \infty$ asymptotic form:
\be
\begin{array}{lll} \label{limH}
\lim_{\t \to \infty}\tilde{\lambda}(\t,\rho,\xh) & = & \lh(\rho,\xh) \\
\lim_{\t \to \infty}\xi_f^a(\t,\rho,\xh)\partial_a & = & \fh(\rho,\xh) \partial_\t \\
\lim_{\t \to \infty}\xi_V^a(\t,\rho,\xh)\partial_a & = & \Vh^\alpha(\rho,\xh) \partial_\alpha .
\end{array}
\ee
Assuming such limiting behaviour at time-like infinity, then Eqns. (\ref{lamo}), (\ref{st}), (\ref{vecf}), (\ref{boxlam}) lead to the following equations on $\H$ \cite{cl3,cl4}:\footnote{By different considerations, the differential equation in  (\ref{eqfh}) was already proposed in \cite{longhi} as a way to define an action of supertranslations on massive fields.}
\ba
\Delta \lh =0 ,  & \quad   \lim_{\rho \to \infty} \lh(\rho,\xh) = \lambda(\xh) , \label{eqlh} \\
\Delta \fh = 3 \fh   ,& \quad \quad  \lim_{\rho \to \infty} \rho^{-1} \fh(\rho,\xh) = f(\xh)  , \label{eqfh} \\
\Delta \Vh^\alpha = 2 \Vh^\alpha, \quad D_\alpha \Vh^\alpha =0, & \quad  \lim_{\rho \to \infty} \Vh^A(\rho,\xh) = V^A(\xh) \label{eqVh},
\ea
where $\Delta$ and $D_\alpha$ are the Laplacian and covariant derivative on $\H$. These boundary-value differential equation problems can be solved by means of Green's functions techniques.  Mathematically the  problems are  equivalent to that of finding certain free fields in Euclidean $AdS_3 \equiv \H$ with prescribed boundary value, for which  the corresponding Green's functions are well known from AdS/CFT literature.

\section{Green's functions on $\H$}\label{greensec}
As before we use $(\rho,\xh)$ as `bulk' coordinates  for $\H$. The sphere boundary of $\H$ will be parametrized by $\qh$. We denote by $\Gu(\rho,\xh;\qh),\Gst(\rho,\xh;\qh)$ and $G^\alpha_B(\rho,\xh;\qh)$ the Green's functions associated to the respective boundary-value problems  (\ref{eqlh}), (\ref{eqfh}) and (\ref{eqVh}). That is, we write the solution to these equations as \cite{cl3,cl4}:
\ba
\lh(\rho,\xh) & = & \int_{S^{2}}d^{2} \qh \, \Gu(\rho,\xh;\qh) \lambda(\qh) , \\
\fh(\rho,\xh)  & =& \int_{S^2} d^2 \qh \, \Gst(\rho,\xh; \qh) f(\qh) , \label{intGf} \\
\Vh^\alpha(\rho,\xh) & =& \int_{S^2} d^2 \qh \, G^\alpha_A(\rho,\xh; \qh) V^A(\qh)  \label{intGV},
\ea
with the Green's functions satisfying:
\be
\Delta  \Gu =0 ,  \hspace{2cm} \lim_{\rho \to \infty} \Gu(\rho,\xh ; \qh) = \delta^{(2)}(\xh,\qh),  \label{greenl} 
\ee
\be
\Delta  \Gst= 3 \Gst ,  \hspace{2cm} \lim_{\rho \to \infty} \rho^{-1} \Gst(\rho,\xh ; \qh) = \delta^{(2)}(\xh,\qh) , \label{greenf} 
\ee
\be
\Delta G^\alpha_B =2 G^\alpha_B , \quad  D_\alpha G^\alpha_B=0,  \quad  \lim_{\rho \to \infty} G^A_B(\rho,\xh; \qh) = \delta^A_B \delta^{(2)}(\xh,\qh),
 \label{greenV}
\ee
where $\Delta$ and $D_\alpha$ act on the $(\rho,\xh)$  variables. In the following sections we describe these Green's functions. It will however be useful to first describe general scalar Green's functions on $\H$, of which $\Gu$ and $\Gst$ are special cases. We will then describe the sphere vector field Green's function $G^\alpha_B$.

\subsection{General scalar Green's functions}
In this section we describe a  family of scalar Green's functions $G^{(n)}(\rho,\xh; \qh)$ parametrized by a real number $n>1$:
\be
G^{(n)}(\rho,\xh; \qh) =  \frac{(n-1)}{ 2^{n-1}} \frac{\sqrt{\gamma(\qh)}}{2 \pi} \big(\sqrt{1+\rho^2}-\rho \, \qh \cdot \xh\big)^{-n} \label{defgn}
\ee
 ($\sqrt{\gamma(\qh)}$  denotes the area element of the boundary sphere). These are  `global coordinates' version of the  free field Green's functions described in \cite{witten,mathur}, specialized to the case of Euclidean $AdS_3 \equiv \H$. Below we show these functions satisfy:
\be
\Delta_{(\rho,\xh)}G^{(n)} = n (n-2) G^{(n)}, \quad  \quad \lim_{\rho \to \infty} \rho^{2-n}G^{(n)}(\rho,\xh; \qh) = \delta^{(2)}(\xh,\qh). \label{Gn}
\ee
For $n=2,3$ Eq.  (\ref{Gn}) becomes  Eq. (\ref{greenl}),   (\ref{greenf}) respectively, so that:
\be
\Gu = G^{(2)}, \quad \Gst =G^{(3)}. \label{G23}
\ee 
We will later  see that  $G^{(4)}$ is relevant for the sphere vector field Green's function $G^\alpha_B$.

To show the first equation in (\ref{Gn}), we `undo' the change of coordinates presented in section \ref{prelsec}  and think of  $ (\rho,\xh) \mapsto G^{(n)}( \rho,\xh ; \qh)$ as a function on (the $t>r$ region of)  Minkowski space that depends only on the hyperboloid direction. In these coordinates, a space-time point $x^\mu$ is parametrized as:
\be
x^\mu = \tau (\sqrt{1+\rho^2}, \rho \xh).
\ee
Let 
\be
q^\mu=(1,\hat{q}) , \label{defq}
\ee
be the future  null 4-vector associated to $\qh$.  Then the function we are interested may be written as:
\be
\big(\sqrt{1+\rho^2}-\rho \, \qh \cdot \xh\big)^{-n} = (q \cdot x/\t)^{-n} \label{Gn2}
\ee
where $q \cdot x = q_\mu x^{\mu}= x^0 - \vec{x} \cdot \qh$. (In this section only we use opposite signature convention to that of section \ref{prelsec} as it simplifies expressions.)  In these coordinates the flat space wave operator takes the form
\be
\square = \partial^2_\tau - \tau^{-2} \Delta ,\label{kg}
\ee
where $\Delta \equiv \Delta_{(\rho,\xh)}$ is the Laplacian in the unit hyperboloid. The action of $\Delta$ on (\ref{Gn2}) may be computed by writing the Laplacian as
\be
\Delta= -\tau^2 \big( \square +\partial^2_\tau \big), 
\ee
and acting on the function in the form given by the RHS of (\ref{Gn2}). Since the action of  $\partial_\tau$ on this function vanishes, we only need to compute the   $\square = \partial^\mu \partial_\mu$ term. Using 
\be
\partial_\mu \tau = \tau^{-1} x_\mu, \quad \partial_\mu (q \cdot x)= q_\mu, \quad q_\mu q^\mu=0, \label{idsxtau}
\ee
one finds
\be
 -\tau^2 \square   ( q \cdot x/\t)^{-n}  = n (n-2)   (q \cdot x/\t)^{-n}, \label{squaregn}
\ee
from which the first equation in (\ref{Gn}) follows. To see the second equation in (\ref{Gn}) we note that the $\rho \to \infty$ behaviour of $G^{(n)}$ depends on whether $\xh$ and $\qh$ coincide or not according to:
\be
\rho^{2-n}G^{(n)}(\rho,\xh; \qh) = \left\{ \begin{array}{lll} O(\rho^{2(1-n)}) & \text{if} & \xh \neq \qh \\ O(\rho^{2}) & \text{if} & \xh = \qh  \end{array} \right. \label{limgn}
\ee
On the other hand, the integral of $G^{(n)}$ over the sphere $\qh$ can be computed and is given by,
 \be
  \int d^2 \qh    \, G^{(n)}(\rho,\xh; \qh)   = \frac{1}{2^{n-1} \rho}\big[ (\sqrt{1+\rho^2}+\rho)^{n-1}-(\sqrt{1+\rho^2}-\rho)^{n-1} \big], \label{intgn}
 \ee
from which it follows that
\be
\lim_{\rho \to \infty} \rho^{2-n} \int d^2 \qh    \, G^{(n)}(\rho,\xh; \qh)  = 1 . \label{limintgn}
\ee
Equations (\ref{limgn}) and (\ref{limintgn}) imply that $\rho^{2-n}  G^{(n)}(\rho,\xh; \qh)$ approaches the delta function as desired. This completes the proof of (\ref{Gn}). 

We close by noting that from Eq. (\ref{G23}) and Eq. (\ref{intgn}) for  $n=2,3$  one finds:
\be
\int d^2 \qh \, \Gu =1, \quad \quad \int d^2 \qh \, \Gst = \sqrt{1+ \rho^2} .\label{intg23}
\ee
The first condition implies that global $U(1)$ transformation at null infinity ($\lambda(\xh)=1$) are indeed mapped to global $U(1)$ transformations at time-like infinity ($\lh(\rho,\xh)=1$). As we will see in section \ref{poincare}, the second condition implies similar agreement regarding time-translations at null and time-like infinity.

\subsection{Sphere vector field Green's function}
The sphere vector field Green's function $G^\alpha_B$ has the tensor structure of a bulk vector and boundary covector. A natural way to incorporate such tensor structure is in terms of bulk and boundary Lorentz generators. In spacetime notation, the bulk Lorentz generators are given by:
\be
J_{\mu \nu} =  x_\mu \partial_\nu - x_\nu \partial_\mu \quad \mu,\nu=0,1,2,3 . \label{defJ}
\ee
Note that although written in spacetime notation, we are regarding  (\ref{defJ}) as a set of 6 vector fields on $\H$ (the explicit form in terms of $(\rho,\xh)$ coordinates is given in Eqns.  (\ref{Jij}), (\ref{Ji0})).
On the other hand, the Lorentz generators on the boundary sphere are given by
\be
L^B_{\mu \nu} = q_\mu D^B q_\nu - q_\mu D^B q_\nu \quad \mu,\nu=0,1,2,3, \label{bdylorentz}
\ee
with $q^\mu$ as in Eq. (\ref{defq}) and $D_B$ the 2-d boundary derivative.  We show below that in terms of these bulk and boundary Lorentz generators, the sphere vector field Green's function is given by:
\be
G^\alpha_B \partial_\alpha = - G^{(4)} L^{\mu \nu}_B J_{\mu \nu} \label{defGV}
\ee
(the index  $B$ is lowered with the metric $\gamma_{AB}$ of the boundary sphere).

We employ the same method of the previous section and realize $G_B \equiv G_B^\alpha \partial_\alpha$ as a vector field in Minkowski space that depends only on the hyperboloid direction.  Up to $x$-independent proportionality factors, the vector field of interest can be written as:
\be
G_{B} \propto (q \cdot x/\t)^{-4}  X^\mu \partial_\mu, \quad  X^\mu =\epsilon^{\mu \nu \sigma \rho} x_\nu q_\sigma \partial_{B} q_\rho, \label{GBX}
\ee
where $\epsilon^{\mu \nu \sigma \rho}$ is the totally antisymmetric symbol. Since $G_B$ is independent of $\t$, one verifies that the 4-d divergence coincides with the 3-d divergence as a vector field on $\H$. From the 4-d perspective it is then immediate to check that the divergence of (\ref{GBX}) vanishes.  The statement that the vector field is annihilated by the differential operator $\Delta-2$ on $\H$ is equivalent to the statement that vector field satisfies the wave equation from the 4-d perspective. The latter can easily be calculated:
\be
\square \big[ (q \cdot x/\t)^{-4}  X^\mu \big] =  X^\mu \square (q \cdot x/\t)^{-4} + 2 \partial_\alpha (q \cdot x/\t)^{-4} \partial^\alpha X^\mu +(q \cdot x/\t)^{-4} \square X^\mu =0,
\ee
where we used Eq. (\ref{squaregn}) for $n=4$, as well as equations (\ref{idsxtau}) and $\square X^\mu=0$.  This concludes the proof of the differential equation identities in (\ref{greenV}).

We now show  the Green's function satisfies the desired boundary condition. To this end, we need to write $J_{\mu \nu}$ in $(\rho,x^A)$ coordinates, and look at the  $\partial_A$ components.  If we introduce the 4-vector
\be
P^\mu:=(\sqrt{1+\rho^2},  \rho \xh),
\ee
the components of interest can be written as (see  Eqns. (\ref{Jij}), (\ref{Ji0})):
\be
J^A_{\mu \nu} = 2 \rho^{-2} P_{[\mu}D^A P_{\nu]}.
\ee
Next, we need to compute the contraction
\be
 L^{\mu \nu}_{B} J^A_{\mu \nu}  = 2 \rho^{-2} \big[  q \cdot P \partial_{B} D^A(q \cdot P) - \partial_{B} (q \cdot P) D^A(q \cdot P) \big].
\ee
To proceed, we choose  $(w,\wb)$ coordinates for $\qh$ and   $(z,\zb)$ coordinates for $\xh$ so that
\ba
\qh & =& (1+w \wb)^{-1} (w+\wb,-i(w-\wb),1-w \wb)  \label{qhwwb}\\
\xh &=& (1+ z \zb)^{-1} ( z+\zb,-i(z-\zb),1-z \zb ).
\ea
For $B= \wb$ one finds
\be
L^{\mu \nu}_{\wb} J^A_{\mu \nu} \partial_A =  \frac{2}{\rho (1+ w \wb)^2}\big[ (\sqrt{1+\rho^2}+\rho) (w-z)^2 \partial_z - (\sqrt{1+\rho^2}-\rho) (1+ w \zb)^2 \partial_{\zb} \big]. \label{LJ}
\ee
The Green's function as defined in Eq. (\ref{defGV}) is obtained by multiplying (\ref{LJ}) with $-G^{(4)}$.  From (\ref{limgn}) and (\ref{LJ}) we see that
\be
G^{z}_{\wb} = \left\{ \begin{array}{lll} O(\rho^{-4}) & \text{if} & \xh \neq \qh \\ 0 & \text{if} & \xh = \qh  \end{array} \right. 
\ee
so that $G^z_{\wb} \to 0$. For the other component we have:
\be
G^{\zb}_{\wb} = G^{(4)}  \times \big[ \rho^{-2} \frac{(1+ w \zb)^2}{(1+ z \zb)^2} +O(\rho^{-3})  \big]\to \delta^{(2)}(w-z),
\ee
where we expanded (\ref{LJ}) in $1/\rho$ and used the limiting condition (\ref{Gn}) for $n=4$. One can repeat the analysis for the the $B=w$ component to find that $G^{\zb}_w \to 0$ and $G^{z}_w \to \delta^{(2)}(w-z)$. This concludes the proof of the boundary condition in (\ref{greenV}).

\section{Green's functions and soft factors}\label{softsec}
The Green's functions described above where encountered in relation with `soft factors'  that appear in the various soft theorems of scattering amplitudes.  In this section we describe such relations.  To this end let us  parametrize  $\H$ by a 3-momentum $\vec{p}$ according to (\ref{prho}) and 
parametrize $\qh$ in terms of $(w,\wb)$ as in Eq. (\ref{qhwwb}). Then, the relations between the Green's functions and the soft factors are \cite{cl3,cl4}:
\ba
 \Gu(\vec{p}/m; w,\wb) & = &   \frac{1}{2\pi}\partial_{\wb} (\sqrt{\gamma_{w \wb}}\frac{\e \cdot p}{ q \cdot p}), \label{id1} \\
\Gst(\vec{p}/m; w,\wb)  &=& - \frac{1}{2\pi  m} \partial_{\wb}\big( \gamma^{w \wb} \partial_{\wb} (\gamma_{w \wb}   \frac{(\e \cdot p)^2}{q \cdot p} ) \big),\label{id2}\\
G^\alpha_{\wb}(\vec{p}/m; w,\wb) \partial_\alpha &= &- \frac{1}{4\pi} \partial^3_{\wb}( \frac{\e \cdot p}{q \cdot p} \, \e^\mu q^\nu ) J_{\mu \nu} .\label{id3}
\ea
Here $\e^{\mu}=1/\sqrt{2}(\wb,1,-i,-\wb)$ is a positive helicity polarization vector associated to $q^\mu \equiv (1,\qh)$,  $\gamma_{w \wb} =2 (1+ w \wb)^{-2}$,  $p^\mu$ the 4-momentum associated to $\vec{p}$ and $J_{\mu \nu}= 2 p_{[\mu}\partial_{\nu]}$ the bulk Lorentz vector fields.\footnote{In this and remaining sections we are back to $(-1,+1,+1,+1)$ signature convention.}

The general structure of the RHS of equations (\ref{id1}), (\ref{id2}), (\ref{id3}) is that of a 2-d boundary differential operator acting on the soft factors that feature in the soft theorems.  The precise form of the 2-d differential operator is dictated by the  `soft charges'  associated to the given symmetry, which in turn can be obtained from phase space methods. 

To show the above relations it is convenient to use a different normalization for the vectors that appear in the soft factors. Define:
\ba
W^\mu & : =& (1+w \wb,w+\wb,-i(w-\wb),1-w \wb) = (1+w \wb) q^\mu  , \label{defW}\\
  E^\mu & :=  &\partial_w W^\mu  = \sqrt{2} \, \e(w,\wb) ,\\
  P^\mu & :=  &  (\sqrt{1+\rho^2},\rho \xh)  = p^\mu/m.
\ea
We will denote the dot product between such vectors by $EP=E_\mu P^\mu$, etc. A key identity to show the relations is:
\be
\partial_{\wb}(EP) WP  - EP \partial_{\wb} (WP) =1, \label{keyeq}
\ee
which can be verified by explicit computation. From (\ref{keyeq}) it immediately follows that:
\be
\partial_{\wb}\frac{EP}{WP} = (WP)^{-2} \label{id11}.
\ee
Expressing  (\ref{id11}) in terms of the original vectors $q,\e,p$, one obtains Eq. (\ref{id1}).

The relevant identity for supertranslations is:
\be
 \partial_{\wb} \big[ (1+ w \wb)^2 \partial_{\wb}  \big( \frac{EP}{(1+ w \wb)} \frac{EP}{WP} \big) \big] = 2 (1+ w \wb) (WP)^{-3} ,\label{Gstapp}
\ee 
which can be shown by repeatedly using  Eqns. (\ref{keyeq}), (\ref{id11}) as well as,
\be
\partial^2_{\wb} W^\mu = \partial^2_{\wb} E^\mu =0. \label{p2w}
\ee
Multiplying (\ref{Gstapp}) by $-(4\pi)^{-1}$ one recovers Eq. (\ref{id2}).

We finally discuss Eq. (\ref{id3}). In terms of the $W,P,E$ vectors it takes the form:
\be
G_{\wb} = -\frac{1}{8 \pi} \partial^3_{\wb} \big[ (\frac{EP}{WP}) EP \,  W^\mu \partial_\mu - EP E^\mu \partial_\mu \big] . \label{greenVapp}
\ee
where $\partial_\mu \equiv \partial/\partial P^\mu$. Since $\partial^2_{\wb} E^\mu=0$, the second term in (\ref{greenVapp}) gives vanishing contribution. Using the identities (\ref{keyeq}), (\ref{id11}), (\ref{p2w}) one finds:
\be
G_{\wb}= -\frac{3}{4 \pi} (WP)^{-4}( WP \partial_{\wb}W^\mu - \partial_{\wb}(WP) W^\mu) \partial_\mu.  \label{gV2}
\ee
The vector field in (\ref{gV2}) can be written as 
\be
(WP \partial_{\wb}W^\mu - \partial_{\wb}(WP) W^\mu) \partial_\mu = W^{[\mu} \partial_{\wb}W^{\nu]} J_{\mu \nu},\label{gv3}
\ee
where 
\be
J_{\mu \nu} = P_\mu \partial_\nu - P_\nu \partial_\mu
\ee
are the bulk Lorentz vector fields.  The  factor multiplying $J_{\mu \nu}$ in (\ref{gv3})  can be seen to be related with the boundary Lorentz generators (\ref{bdylorentz}) by
\be
   W^{[\mu} \partial_{\wb} W^{\nu]} = \frac{(1+w \wb)^2}{2} L^{\mu \nu}_{\wb}  \label{LW}
\ee
Substituting these expressions in (\ref{gV2}) and writing everything in terms of the original vectors $q,\e,p$, one recovers Eq. (\ref{id3}).

\section{Poincare subgroup of generalized BMS group} \label{poincare} 
Let $x^\mu=(t, \vec{x})$ be Cartesian coordinates of Minkowski spacetime. The spacetime Poincare generators are given by the vector fields: 
\be
\partial_\mu, \quad J_{\mu \nu} := x_\mu \partial_\nu -x_\nu \partial_\mu , \quad \quad \mu,\nu=0,1,2,3. \label{poincgenmin}
\ee
By going to outgoing null coordinates $r,u:=t-r,\xh$ and taking  $r \to \infty$, one finds that the corresponding  generators  at null infinity are:
\be
\begin{array}{llll} \label{poincG}
T_0(\qh) & := &1 \quad & \text{time translations}  \\
T_i(\qh) &:= & -\qh_i  \quad & \text{space translations} \\
L^A_{ij}(\qh) &: = &\qh_i D^A \qh_j - \qh_j D^A \qh_i \quad & \text{rotations} \\
L^A_{i0}(\qh) &: =& D^A \qh_i  \quad  & \text{boosts} 
\end{array}
\ee
These are the generators of the Poincare subgroup  inside the  generalized BMS group of Eqns. (\ref{st}), (\ref{vecf}).\footnote{In the present case there is a preferred Poincare subgroup associated to the Minkowski metric. In a generic asymptotically flat spacetime there is no unique asymptotic Poincare subgroup \cite{aareview}.}

 On the other hand, expressing the vector fields (\ref{poincgenmin}) in the hyperbolic coordinates of section (\ref{prelsec}) one finds: 
\ba
\partial_0 & = & \sqrt{1+ \rho^2} \, \partial_\t +O(\t^{-1}) \label{partial0}\\
\partial_i  & = & - \rho \, \xh_i \, \partial_\t +O(\t^{-1}) \label{partiali}
\ea
\ba
J_{i j}  & = &  (\xh_i D^A \xh_j - \xh_j D^A \xh_i)\partial_A \label{Jij} \\
J_{i 0} & = & \sqrt{1+ \rho^2}( \rho^{-1} D^A \xh_i \partial_A + \xh_i \partial_\rho). \label{Ji0}
\ea
We now verify that the time-like asymptotic value of the Poincare vector fields (\ref{partial0}), (\ref{partiali}), (\ref{Jij}), (\ref{Ji0}) 
coincide with what is obtained from the Green's function formulas (\ref{intGf}), (\ref{intGV}) applied to the boundary generators (\ref{poincG}). 


\subsection{Spacetime translations}
We recall that the supertranslation Green's functions is given by Eq. (\ref{defgn}) for $n=3$. Then, according to (\ref{intGf}) the asymptotic $\t$ component of the spacetime translations (\ref{partial0}), (\ref{partiali}) is given by:
\be
\xio^\t(\rho,\xh) = \int d^2 \qh \,  G^{(3)}(\rho,\xh; \qh)T_\mu(\qh). \label{fhapp}
\ee
For time translations $T_0(\qh)=1$ the integral is given by  Eq. (\ref{intgn}) for $n=3$ which becomes:
\be
\int d^2 \qh    \, G^{(3)}  = \sqrt{1+\rho^2} \label{intg3}
\ee
in agreement  with (\ref{partial0}).  In order to evaluate the integral (\ref{fhapp}) for spatial translations $T_i(\qh)=- \qh_i$ we express the integrand in terms of derivatives of  $G^{(2)}$ as follows.  

Let $\vec{P}= \rho \xh$ so that $\rho= \sqrt{\vec{P} \cdot \vec{P}}$. The integrand we are interested can then be written as:
\be
 G^{(3)}(\vec{P} ; \qh) \qh_i = \frac{1}{2} \frac{\partial}{\partial P_i} G^{(2)}(\vec{P}; \qh) + \frac{P_i}{\sqrt{1+\rho^2}} G^{(3)}(\vec{P}; \qh). \label{partialG2}
\ee
Integrating over $\qh$ and using Eq. (\ref{intgn}) for $n=2$ (which gives 1 and hence gives zero contribution in (\ref{partialG2})) and Eq. (\ref{intg3}) one obtains
\be
\int d^2 \qh \, G^{(3)}(\rho,\xh; \qh) \qh_i = \rho \xh_i,  \label{intg3qi}
\ee
 in agreement with (\ref{partiali}).   
\subsection{Boosts}
Using the form of the Green's function given in (\ref{defGV}), the integral formula (\ref{intGV}) for the hyperboloid vector field  associated to the boost $V^B(\qh)=L^B_{i0}$    takes the form:
\be
\xio^\alpha(\rho,\xh)  = -\int d^2 \qh \, G^{(4)}(\rho,\xh; \qh) L^B_{i0}(\qh) L_{B}^{\mu \nu}(\qh) J^\alpha_{\mu \nu} ,\label{Vhapp}
\ee
where $J^\alpha_{\mu \nu} \partial_\alpha$ is given by Eqns. (\ref{Jij}), (\ref{Ji0}) and $L^{\mu \nu}_B(\qh)$ by Eq. (\ref{bdylorentz}) (or equivalently Eq. (\ref{poincG})).  Note that  $J^\alpha_{\mu \nu} \partial_\alpha$ is independent of $\qh$ so that it can be taken  outside the integral.  The contraction $L^B_{i0}(\qh) L_{B}^{\mu \nu}$ can be evaluated with help of the identity:
\be
D^A \qh_i D_A \qh_j= \delta_{ij}- \qh_i \qh_j
\ee
and gives
\ba
L^B_{i0}(\qh)   L_{B}^{j 0}(\qh) & =& \qh_i \qh^j - \delta^j_i   \label{Lj0} \\
L^B_{i0}(\qh)   L_{B}^{j k}(\qh) & =&  \qh^j \delta^k_i-\qh^k \label{Ljk} \delta^j_i.
\ea
In order to evaluate (\ref{Vhapp}) we then need  the integrals of $G^{(4)}$, $ \qh_i G^{(4)}$ and $\qh_i \qh_j G^{(4)}$. The first one is given in Eq. (\ref{intgn}) for $n=4$ and reads:
\be
\int d^2 \qh    \, G^{(4)}  = 3/4 + \rho^2.   \label{intg4}
\ee
 For the two other ones we use similar trick as in Eq. (\ref{partialG2}). Taking derivative of $G^{(3)}$ one finds:
\be
G^{(4)}(\vec{P} ; \qh) \qh_i =\frac{1}{4} \frac{\partial}{\partial P_i} G^{(3)}(\vec{P}; \qh) + \frac{P_i}{\sqrt{1+\rho^2}} G^{(4)}(\vec{P}; \qh). \label{partialG3}
\ee
Integrating (\ref{partialG3}) over the sphere and using (\ref{intg3}), (\ref{intg4}) one gets:
\be
\int d^2 \qh \, G^{(4)}( \rho, \xh ; \qh) \qh_i  = \rho \sqrt{1+\rho^2} \, \xh_i \label{intg4qi}
\ee
Finally, multiplying (\ref{partialG3}) by  $\qh_j$ and using Eqns. (\ref{intg3qi}), (\ref{intg4qi}) one finds:
\be
\int d^2 \qh \, G^{(4)}(\rho,\xh ; \qh) \qh_i \qh_j = \frac{1}{4} \delta_{ij} + \rho^2 \xh_i \xh_j .\label{intg4qiqj}
\ee
Collecting all results in (\ref{Vhapp}) one arrives at:
\be
\xio^\alpha =  J^\alpha_{i 0} - 2 \rho^2 [\xh_i \xh^k J^\alpha_{k 0} - J^\alpha_{i 0} + \rho^{-1}\sqrt{1+\rho^2} \xh^k J_{k i}] .\label{xioboost}
\ee
Using the expressions for $J^\alpha_{\mu \nu}$ given in (\ref{Jij}), (\ref{Ji0}) one can verify that the term in the square bracket vanishes an we recover the boost vector field  (\ref{Ji0}). 

\subsection{Rotations}
Since the verification of rotations goes along the same lines as for boosts we give a summarized version.  The boundary vector field in (\ref{Vhapp}) is now taken to be $L^B_{ij}(\qh)$ (in place of $L^B_{i0}(\qh)$). 
The contractions are now:
\ba
L^B_{ij}(\qh)  L_{B}^{k 0}(\qh) & =& -2 \qh_{[i} \delta^k_{j]}   \\
L^B_{ij}(\qh)  L_{B}^{k l}(\qh) & =&  -2 \qh_{[i} \delta^k_{j]} \qh^l -  k \leftrightarrow l .
\ea
The relevant integrals can be evaluated from (\ref{intg4qi}), (\ref{intg4qiqj}). Using identities such as $\xh^k \xh_k=1$, $\xh^k D_A \xh_k=0$ and the vanishing of the square bracket in (\ref{xioboost}) one obtains $\xio^\alpha= J^\alpha_{i j}$.\\

\noindent {\bf Acknowledgements} \\
I am  deeply grateful to Alok Laddha for collaboration, discussions,  and for encouraging me to write this note.

\end{document}